\documentclass[]{spie}  

\usepackage{multirow}
\usepackage{floatrow}
\usepackage{amsmath,amsfonts,amssymb}
\usepackage{graphicx}
\usepackage[colorlinks=true, allcolors=blue]{hyperref}

\title{Towards A Device-Independent Deep Learning Approach for the Automated Segmentation of Sonographic Fetal Brain Structures: A Multi-Center and Multi-Device Validation}

\author[a]{Abhi Lad}
\author[a]{Adithya Narayan}
\author[a]{Hari Shankar}
\author[b]{Shefali Jain}
\author[c]{Pooja Punjani Vyas}
\author[d]{Divya Singh}
\author[e]{Nivedita Hegde}
\author[a]{Jagruthi Atada}
\author[f]{Jens Thang}
\author[g]{Saw Shier Nee}
\author[h]{Arunkumar Govindarajan}
\author[e]{Roopa PS}
\author[e]{Muralidhar V Pai}
\author[e $\star$]{Akhila Vasudeva}
\author[b $\star$]{Prathima Radhakrishnan}
\author[f $\star$]{Sripad Krishna Devalla}
\affil[a]{Origin Health India, Bengaluru, Karnataka, India}
\affil[b]{Bangalore Fetal Medicine Center, Bengaluru, Karnataka, India}
\affil[c]{Jaslok Hospital |\& Research Center, Mumbai, Maharashtra, India}
\affil[d]{Prime Imaging and Prenatal Diagnostics, Chandigarh, Punjab, India}
\affil[e]{Kasturba Medical College, Manipal Academy of Higher Education, Manipal, Karnataka, India}
\affil[f]{Origin Health, Singapore}
\affil[g]{Department of Artificial Intelligence, Faculty of Computer Science and Information Technology, Universiti  Malaya, Kuala Lumpur, Malaysia}
\affil[h]{Aarthi Scans and Labs, Chennai, Tamil Nadu, India}
\affil[$\star$]{Corresponding Authors}

\begin{document} 
\maketitle

\begin{abstract}
Quality assessment of prenatal ultrasonography is essential for the screening of fetal central nervous system (CNS) anomalies. The interpretation of fetal brain structures is highly subjective, expertise-driven, and requires years of training experience, limiting quality prenatal care for all pregnant mothers. With recent advancement in Artificial Intelligence (AI), computer assisted diagnosis has shown promising results, being able to provide expert level diagnosis in matter of seconds and therefore has the potential to improve access to quality and standardized care for all. Specifically, with advent of deep learning (DL), assistance in precise anatomy identification through semantic  segmentation essential for the reliable assessment of growth and neurodevelopment, and detection of structural abnormalities  have been proposed. However, existing works only identify certain structures (e.g., cavum septum pellucidum [CSP], lateral ventricles [LV], cerebellum ) from either of the axial views (transventricular [TV], transcerebellar [TC]), limiting the scope for a thorough anatomical assessment as per practice guidelines necessary for the screening of CNS anomalies. Further, existing works do not analyze the generalizability of these DL algorithms across images from multiple ultrasound devices and centers, thus, limiting their real-world clinical impact. In this study, we propose a deep learning (DL) based segmentation framework for the automated segmentation of 10 key fetal brain structures from 2 axial planes from fetal brain USG images (2D). We developed a custom U-Net variant that uses inceptionv4 block as a feature extractor and leverages custom domain-specific data augmentation. Quantitatively, the mean (10 structures; test sets 1/2/3/4) Dice-coefficients were: 0.827, 0.802, 0.731, 0.783. Irrespective of the USG device/center, the DL segmentations were qualitatively comparable to their manual segmentations. The proposed DL system offered a promising and generalizable performance (multi-centers, multi-device) and also presents evidence in support of device-induced variation in image quality (a challenge to generalizibility) by using UMAP analysis. Its clinical translation can assist a wide range of users across settings to deliver standardized and quality prenatal examinations. 
\end{abstract}

\keywords{fetal ultrasonography, transventricular, transcerebellar, U-Net variant, segmentation, multi-center, multi-device }

\section{Introduction}
\label{sec:intro}  

The prenatal ultrasonography (USG) examination of the fetal brain at the level of the transventricular (TV) and transcerebellar (TC) planes (axial views) during the 2nd trimester has been clinically recommended for the screening of central nervous system (CNS) anomalies \cite{salomon2011practice}. It offers scope for timely specialist referrals, necessary intervention and appropriate pregnancy management. The central nervous system (CNS) anomalies are among the most common types of fetal anomalies (1.4-1.6 per 1000 live births) \cite{onkar2014evaluation}. With 3-6\% stillbirths, low 5-year survival rates, life-long physical and mental disabilities, CNS anomalies take a very expensive toll on families, community, and the healthcare system at large \cite{groen2017stillbirth,schneuer2019five}. Due to poor specialist-to-patient ratio, access to quality and timely prenatal care is limited for pregnant women; especially in low-resource, rural, and semi-urban settings. Several existing deep learning (DL) approaches have focused on only certain structures on any one of the axial views (TC or TV). Some applications include but are not limited to, detecting standard fetal planes and structures, segmentation of fetal anatomy (e.g., brain, head, face, abdomen, cardiac, lungs, etc.), and classification (normal/abnormal/pathology) of fetal USG images \cite{burgos2019evaluation,garcia2020machine,sinclair2018human,sobhaninia2020localization,van2018automated}. Specifically, for USG images of the fetal brain, studies have attempted to individually segment key structures such as the cranium, cavum septum pellucidum (CSP), LV, etc. \cite{chen2020automatic,sinclair2018human,wu2020automatic}. However, these studies have not explored generalizability of these algorithms on images from unseen devices and multiple imaging centers, thus limiting the clinical translation of these approaches. In this study, we propose a custom DL framework for the automated segmentation of 10 key fetal structures across both the TC and TV planes. Further, we validate the generalizability of the proposed approach across 4 independent test sets comprising of images from 4 unseen USG devices across 3 imaging centers. We have also provided UMAP \cite{mcinnes2018umap} analysis of images from different USG machines from across the imaging centers to support the machine induced variance in the USG image characteristics.
Following are the key contributions of this study:
\begin{itemize}
  \item Development, validation, and benchmarking of a U-Net inspired multi-plane, multi-structure DL segmentation model.
  \item UMAP analysis of multi-device USG images to establish image characteristic variance among USG devices.
  \item Significance of domain-specific data augmentation and its effect on DL model generalizability.
  \item Extensive (1626 images) multi-device, multi-center validation to assess the robustness of the proposed approach.
\end{itemize}

\section{MATERIALS AND METHODS}

\subsection{Fetal Ultrasound Imaging}
\label{sec:title}

A total of 4370 two-dimensional (2D) standard plane USG images of the TV and TC planes were obtained from 3 centers (2 tertiary referral centers [TRC 1,2] + 1 routine imaging center [RIC]) using 6 USG devices (General Electric [GE] Voluson: P8, P6, E8, E10, S10; Samsung: HERA W10). The images were acquired from a cohort of 1349 pregnant women who received their targeted mid-trimester USG examination (transabdominal). All USG examinations were reviewed and approved by fetal medicine specialists for quality and correctness. Only live singleton fetuses that did not exhibit any growth anomalies were included in this study. Due to the study's retrospective nature, all informed consent was waived. All data was anonymized as per the tenets of the Declaration of Helsinki after approval from the ethics committee of the respective centers. The USG examination was performed based on internationally recommended practice guidelines \cite{salomon2011practice}.

\subsection{Dataset Preparation}

From the 4370 images, a total of 2744 images (TC/TV images from GE E8: 641/1307; TC/TV images from GE S10: 221/575)  from TRC 1 were used for training and validation of the DL model. Data balancing (up-sampling) was done to ensure equal representation from each USG device and care was taken to ensure there were no duplicate images, or images from the same subjects between the training and validation datasets (training and validation splits: 89.7\%:10.3\%).

The remaining 1626 images were used as a part of 4 test sets. Test set 1 (TRC 1; trained device: GE E8 [82.3 \%], GE S10 [17.7 \%]) consisting of of 719 images (TC/TV: 131/588) was used to test the performance of best DL model. The robustness and generalizability of the DL model was assessed on independent test sets 2,3 and 4. Test set 2 (TRC 1, unseen devices: Samsung HERA W10 [79.2 \%], GE P6 [6.8 \%], GE E10 [14 \%]) consisted of 192 images (TC/TV: 90/102), test set 3 (TRC 2, trained devices: GE E8 [69.5 \%], GE S10 [30.5 \%]) consisted of 557 images (TC/TV: 223/334); and test set 4 (RIC; unseen device: GE P8) consisted of 158 images (TC/TV: 100/58)

All images were resized to 160 (height) x 288 (width). To preserve the inherent variability (i.e., gain, zoom, contrast, speckle noise, patient-specific probe settings, pixel resolution, etc.) in the dataset, no additional pre-processing was done.

Manual segmentations were prepared for all the 4370 images by well-trained medical image annotators and reviewed by fetal medicine specialists. A total of 10 unique fetal brain structures/classes from the TV  (cranium, brain parenchyma, cavum septum pellucidum [CSP], midline falx, choroid plexus, lateral ventricles [LV]) and TC planes (cranium, brain parenchyma, CSP, midline falx, cerebellum, cisterna magna, nuchal fold, skin) were segmented. Note that each class common to both the planes (i.e., cranium, brain parenchyma, midline falx and CSP) were assigned the same unique index while preparing the segmentation masks. 

   \begin{figure} [ht]
   \begin{center}
   \begin{tabular}{c} 
   \includegraphics[width=1\textwidth]{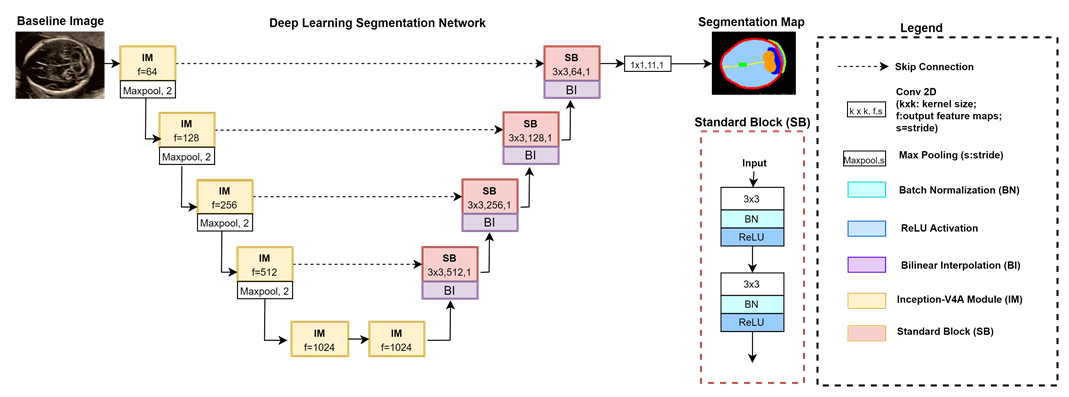}
   \end{tabular}
   \end{center}
   \caption{ \label{fig:architecture_figure} 
The U-Net inspired DL architecture used in this study is shown.}
   \end{figure} 
   
\subsection{Deep Learning Network}

We propose a fully-convolutional and end-to-end DL network inspired (\textbf{Figure \ref{fig:architecture_figure}})by U-Net \cite{ronneberger2015u} to simultaneously segment 10 fetal brain structures across 2 axial views. Inception v4A \cite{szegedy2016inceptionv4} blocks were used as feature extractors for the 4-levels of the (feature maps: 64, 128, 256, 512) encoder and latent space, while standard blocks (\textbf{Figure \ref{fig:architecture_figure}}) were used in the decoder for feature extraction. Segmentation of fetal brain structures can often be difficult especially in the cases of structural anomalies due to common co-existence of multiple fetal anomalies. Specifically for finer structures like LV, CSP, and nuchal fold where subtle changes in structure are associated with critical anomalies like ventriculomegaly, agenesis of corpus callosum, and Down syndrome. The inceptionv4A block with its multiple kernel sizes can take into account the global features, like structures in vicinity e.g. choroid plexus inside LV and local features, like boundaries and integrity of structures e.g. case of choroid plexus with a cyst inside it. The input image (160 [height] x 288 [width]) was downsampled by a factor of 2 at each level of the encoder (maxpooling, stride = 2), and sequentially upsampled to original scale (bi-linear interpolation) in the decoder. The output from the decoder was passed through a 1x1 convolutional layer (number of feature maps= [11; 10 fetal structures + background], stride = 1) followed by softmax activation to obtain the class-wise probability distributions. To get the final segmentation mask (herein referred to as DL segmentation), each pixel was assigned the class with maximum probability.

The entire network was trained end-to-end with Adam \cite{kingma2014adam} optimizer, Dice loss, fixed learning rate of 0.0001, and a mini batch of 2 images. The model with the best validation loss was finally chosen.  The proposed network consisted of 38 million trainable parameters and was trained and tested on an NVIDIA Tesla T4 with CUDA v11.0 and cuDNN v7.6.5 using PyTorch 1.7.0. 

\subsection{Domain-Specific Data Augmentation}

\begin{figure}[h]
\begin{center}
\begin{tabular}{c} 
\includegraphics[width=1\textwidth]{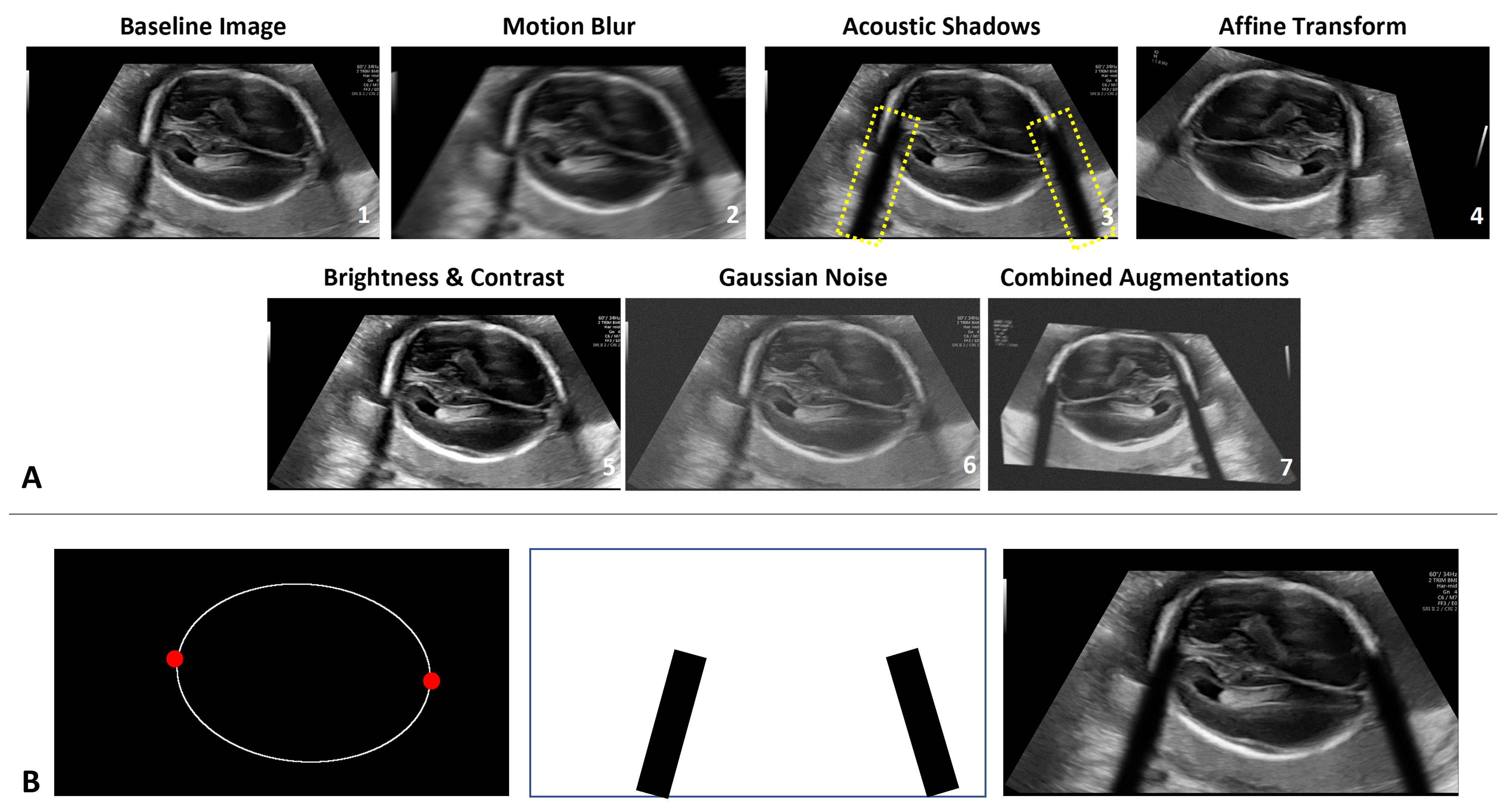}
\end{tabular}
\end{center}
\caption{\label{fig:dataaugmentations} 
 (A) Domain-specific data augmentation (1: baseline; 2: motion blur; 3: acoustic shadows [highlighted in yellow box]; 4: affine transforms; 5: brightness and contrast; 6: Gaussian multiplicative noise; and 7: resultant image of combined augmentations) are shown. (B) The visualization of steps used to create artificial cranium shadows as part of data augmentation from left to right: cranium mask ground truth used to identify left and right extremities; black patches created from extreme points; superimposing black patches on correspoonding USG image to create cranium shadows.}
\end{figure}

Fetal USG images from different USG devices and centers may appear visually different in terms of image quality (e.g., noise, resolution, acoustic shadowing), perception (e.g., brightness, contrast, intensity homogeneity) and presentation (e.g., zoom, orientation; operator dependent factors). To counter these effects, we used extensive domain-specific online data augmentation (DA; implemented using Albumentations \cite{info11020125}; \textbf{Figure \ref{fig:dataaugmentations} A}) to improve the DL model's performance and generalizability across images from different USG devices/centers.

Image quality and perceptual factors were augmented using color-jitter (brightness: 0.8-1.2 and contrast: 0.8-1.2) - representative of various image enhancing settings, quality reduction (downsampling and upsampling the images; range: 0 to 0.5) - representative of poor quality hardware or image compression loss, speckle noise (Gaussian multiplicative noise; mean: 0, standard deviation: 0.1) - represents artifact introduced by proprietary post-processing, and acoustic shadowing - characteristic artifact formed due to acoustic property of cranium.

In this study, we propose a novel approach to accurately represent fetal acoustic shadows occurring due to attenuation of sound beam at the edges of cranium due to incorrect focal depth. As shown in \textbf{Figure \ref{fig:dataaugmentations} B}, the acoustic shadows are created using the cranium segmentation mask which is used to determine horizontal extremities (red dots) of the cranium. Then 2 patches of width 75-100 pixels are extended from the extreme points at angle 15-20 degrees. These patches are further processed with blur to achieve more realistic murky shadows. Then it is superimposed on original image resulting in image with artificial acoustic cranium shadows as shown in \textbf{Figure \ref{fig:dataaugmentations} B}.

We used affine transformations (zoom [range: 0.6 to 1.2], rotation [range: +20 to -20 degrees], translations [range: 40 to 60 pixels along height or width], shear transformations [range: 0 to 0.2]) and motion blur (kernel size: 50x50 pixels) to simulate operator variability. Finally, variation in fetal anatomy due to anomalies/gestation age were accounted by using elastic deformations (standard deviation: 50, affine: 50).

\subsection{Experiments and Analysis}

\subsubsection{UMAP Analysis}

\begin{figure}[h]
\begin{center}
\begin{tabular}{c} 
\includegraphics[width=1\textwidth]{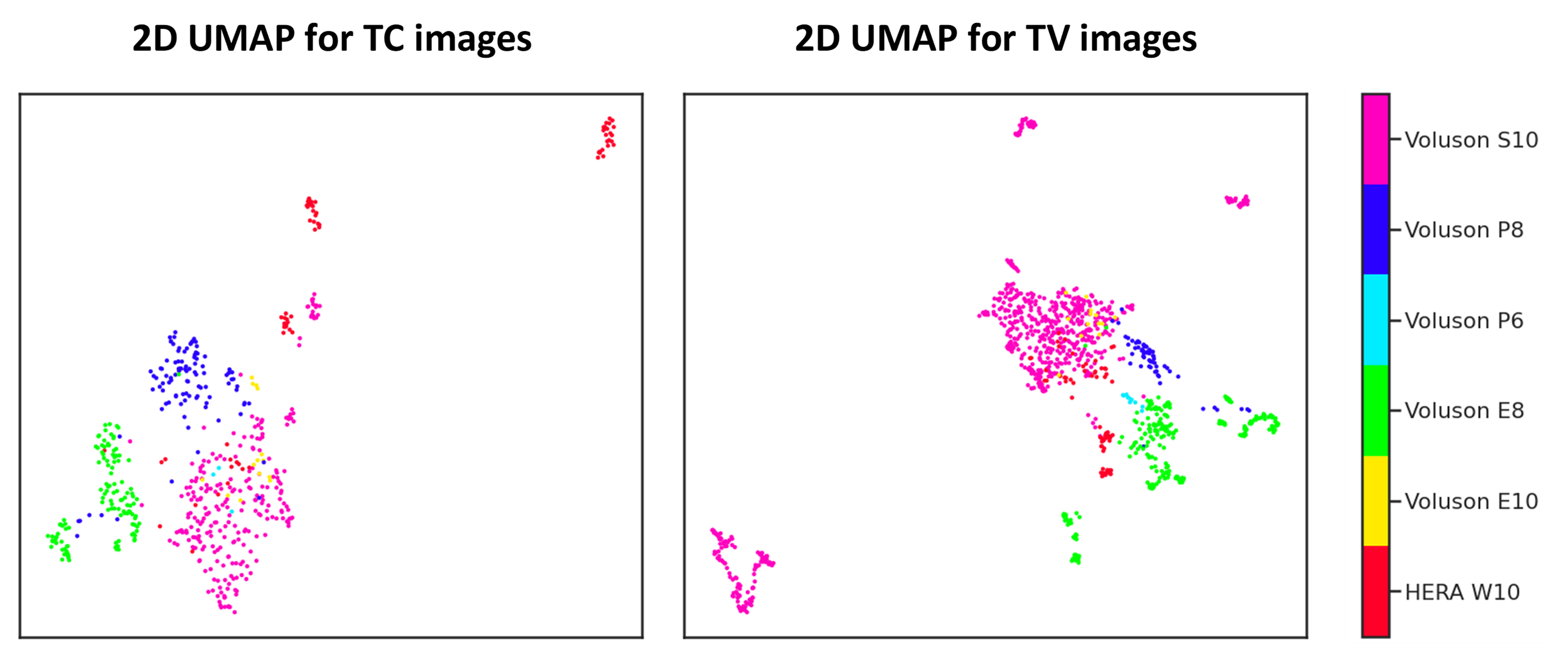}
\end{tabular}
\end{center}
\caption{\label{fig:umap}
2D UMAP (low dimensional) representation of TC images (left) and TV images (right) corresponding to 6 USG devices across all 4 test sets. Overlapping clusters suggest similarity in image characteristics among USG devices. Dense clusters indicate low intra-device variance. Separation of clusters as observed above supports the variance among images from different USG devices is substantial and qualitatively verifiable.
}
\end{figure}

The visual feel/quality of images generated by USG devices vary depending on model, manufacturer and provided operator settings. Such variations in image quality occur due to proprietary image construction/processing algorithms, image compression techniques or even probe technology used. In order to demonstrate the USG device induced variance in images, we use UMAP \cite{mcinnes2018umap} algorithm. The UMAP is a dimensionality reduction technique similar to tSNE which is useful in visualizing high dimensional data in low dimensional (2D,3D) euclidean space. UMAP algorithm creates a connected graph based on high-dimension input data, i.e. images, and projects it to low dimensional (2D, 3D) graph by using force-directed graph layout algorithm. This low dimensional graph is mathematically similar to the high dimensional graph, i.e. the points appearing closer in lower dimensional space are actually similar even in high dimensional space. The presentation of projected low-dimensional graph depends on parameters like nearest-neighbour and minimum distance used to create connected graph (manifold). Separated clusters indicate that images from different USG devices are varied and dense clusters indicate low intra-device variance. Overlapping clusters suggest similarity in image characteristics among USG devices. The resultant plot as presented in \textbf{Figure \ref{fig:umap}} clearly depicts formation of clusters for each USG device. The plots of TC and TV images are presented individually to prevent cluster formation due to different planes. While some overlap is observed among the clusters, specifically for Voluson S10 (\textbf{Figure \ref{fig:umap}}: pink) and P6 (\textbf{Figure \ref{fig:umap}}: cyan), overall the images have enough variance to form a cluster with images from same machine lying in vicinity. Interestingly, images from Samsung Hera W10 (\textbf{Figure \ref{fig:umap}}: red) have high intra-device variance thus resulting in separate clusters for the same USG device.

\begin{figure}[h]
\begin{center}
\begin{tabular}{c} 
\includegraphics[width=1\textwidth]{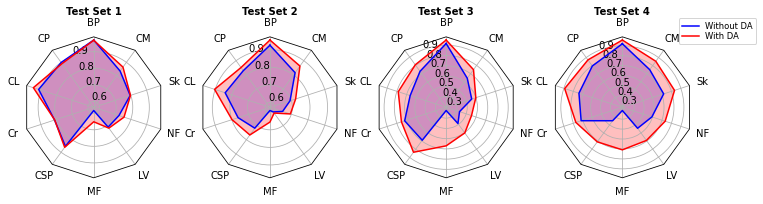}
\end{tabular}
\end{center}
\caption{\label{fig:radialplot} 
The Dice-coefficients (when tested on model trained with and w/o domain-specific DA) for individual structures across the 4 test sets are shown.}
\end{figure} 

\begin{figure}[h]
\begin{center}
\begin{tabular}{c} 
\includegraphics[width=1\textwidth]{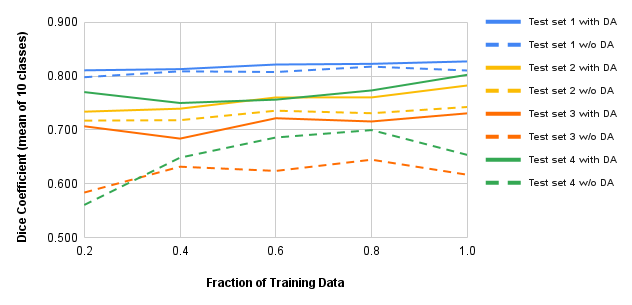}
\end{tabular}
\end{center}
\caption{\label{fig:datasizeablation} 
The effect of training dataset size on the segmentation performance (when tested on model trained with and w/o domain-specific DA) across the test sets is shown.}
\end{figure} 

\begin{figure}[h]
\begin{center}
\begin{tabular}{c} 
\includegraphics[width=1\textwidth]{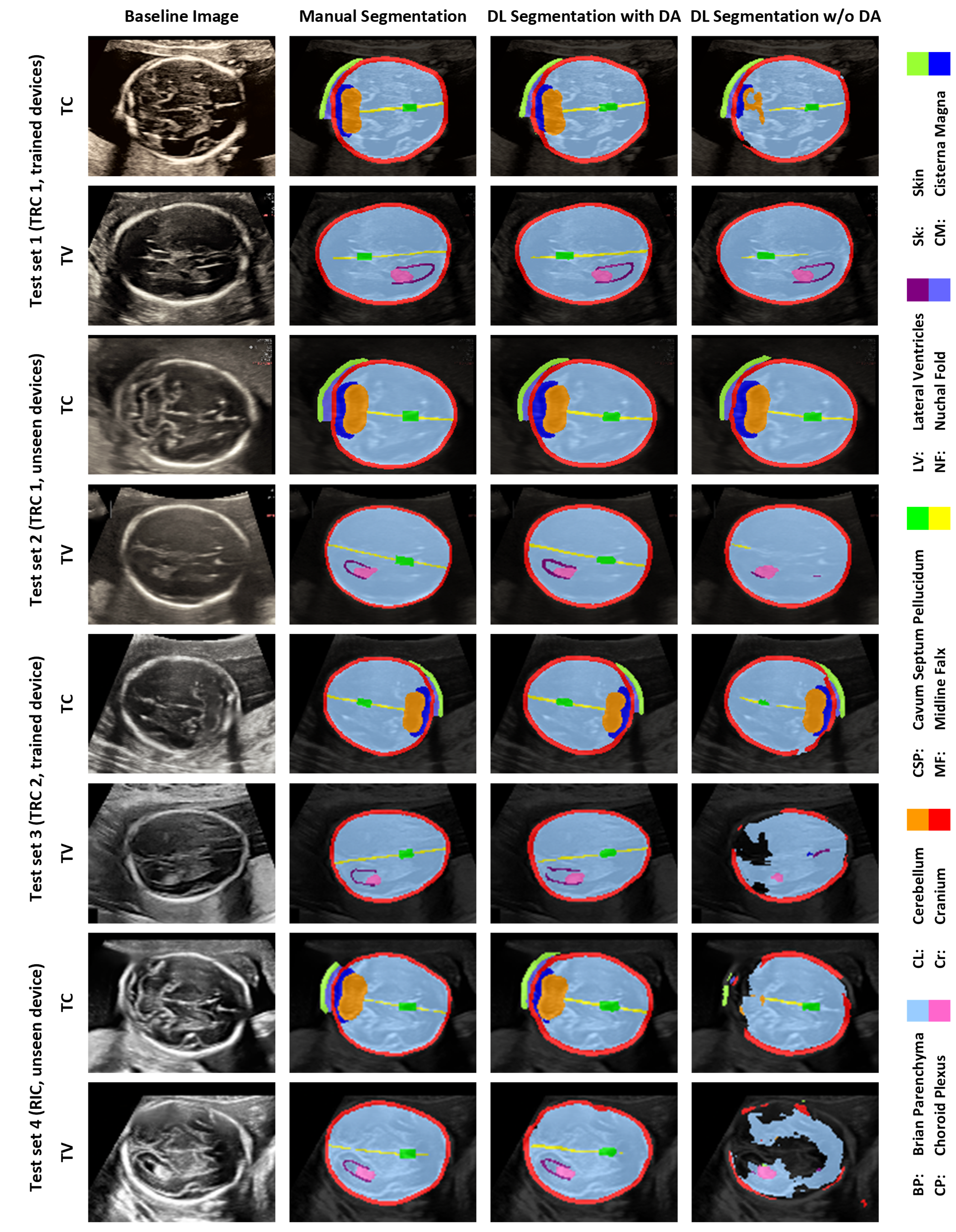}
\end{tabular}
\end{center}
\caption{\label{fig:quality} 
Qualitative comparison of the proposed DL segmentation with and without (w/o) domain-specific DA (vs. manual segmentation). 1st row represents baseline images from 8 subjects (2 per test set), 2nd, 3rd, and 4th columns represents their manual segmentation, and DL segmentations (with and w/o domain specific DA), respectively.}
\end{figure} 

\subsubsection{Deep Learning Experiments and Ablations}

We performed the following experiments: (1) benchmarking against multiple DL segmentation models that have been widely used for segmentation problems (U-Net \cite{ronneberger2015u}, LinkNet \cite{chaurasia2017linknet}, MNet \cite{7950555}, Deeplabv3+ \cite{chen2018encoder} and HRNet \cite{wang2020deep}); (2) effect of domain-specific DA (trained with and without (w/o) domain-specific DA); (3) effect of dataset size (data ablation on fractions [0.2, 0.4, 0.6, 0.8, 1.0] of the training dataset; performed with and without domain-specific DA) and model performance; and (4) 5-fold cross-validation to assess repeatability. 

For all experiments, the DL segmentations from all the 4 test sets were qualitatively and quantitatively (structure wise Dice coefficient [DC] as per standard definition) assessed by benchmarking against the expert-annotated manual segmentations. Note that background class was excluded from qualitative and quantitative comparison for all experiments.

\section{RESULTS AND DISCUSSION}
\label{sec:sections}

\subsection{Qualitative Comparison with Manual Segmentation}
Qualitatively, the DL segmentations were comparable to the manual segmentations across all test sets for both the TV and TC plane structures. As it is evident from \textbf{Figure \ref{fig:quality}}, there is substantial difference between DL segmentations with DA and w/o DA. While DL segmentations with DA (3rd column) are comparable to manual segmentations, the DL segmentations w/o DA (4th column) even fail to correctly segment brain parenchyma, which covers more than half of segmentable area inside the brain region (\textbf{Figure \ref{fig:quality}}: light blue) and thus is essentially unable to predict finer structures such as CSP and LV.

\subsection{Quantitative Comparison of Models}
Quantitatively (\textbf {Table~\ref{table1}}; mean of 5-fold cross validation was used), when benchmarked against widely used segmentation models such as U-Net \cite{ronneberger2015u}, LinkNet \cite{chaurasia2017linknet}, MNet \cite{7950555}, Deeplabv3+, \cite{chen2018encoder} and HRNet \cite{wang2020deep}, we obtained a performance improvement (mean DC) of 0.5\% to 2\%  across 4 test sets. Specifically, compared to the benchmarking models, our model localized the the finer structures (LV, CSP, nuchal fold) that are crucial in the screening of CNS anomalies and overall resulted in better performance, i.e. average improvement over benchmarking models for CSP: 2.4 \%, LV: 9.8 \% and nuchal fold: 8.3 \% when averaged across test sets.

\subsection{Effect of Domain Specific Data Augmentations}
 The effect of domain-specific DA on the DL segmentations were qualitatively (\textbf{Figure \ref{fig:quality}}, 3rd and 4th columns) and quantitatively observed (\textbf{Figure \ref{fig:radialplot}}). As it is evident from (\textbf{Figure \ref{fig:quality}}) columns 3rd and 4th, data augmentation can have a profound impact on the segmentation performance especially for USG images from different centers. Across all test sets (especially unseen devices/unseen centers), in general, finer structures and structures of critical importance like LV, CSP and cerebellum were affected the most (\textbf{Figure \ref{fig:quality}}, 4th column) in the absence of domain-specific DA. Specifically, we observed performance gains (percentage change in mean DC; \textbf{Figure \ref{fig:radialplot}} and \textbf {Table~\ref{table1}}) of 2.1\%, 7.11\%, 18.4\%, and 22.7\% across test sets 1, 2, 3, and 4 respectively. Besides, even when the domain-specific DA were used with the benchmarking models, on an average, we observed a minimum of 0.486\% and a maximum of 30\% performance gain for all cases.

\subsection{Ablation on Size of Training Set}
When trained (with domain-specific DA) on increasing fractions of dataset (0.2, 0.4, 0.6, 0.8, 1.0), across all test sets, the DL segmentation performance continued to improve (\textbf{Figure \ref{fig:datasizeablation}}). When repeated the same without domain-specific DA, for a given fraction of training data, the segmentation performance was overall (mean of all fractions; vs. trained with domain-specific DA) reduced by 1\%, 4\%, 11.4\%, and 14.8\%, across test sets 1, 2, 3 and 4 respectively.

\begin{table}[h]
\centering
\caption{Benchmarking of the proposed DL method (with and without domain-specific DA) against U-Net~\cite{ronneberger2015u}, LinkNet~\cite{chaurasia2017linknet}, MNet~\cite{7950555}, DeepLabv3+~\cite{chen2018encoder}, and HRNet~\cite{wang2020deep} across 4 independent test sets. Mean Dice coefficient (DC; mean of all 10 classes) was used for the purposes of evaluation.}
\begin{tabular}{|l|c|c|c|c|c|c|c|c|} 
\hline
\multirow{2}{*}{Models} & \multicolumn{2}{c|}{Test set 1} & \multicolumn{2}{c|}{Test set 2} & \multicolumn{2}{c|}{Test set 3} & \multicolumn{2}{c|}{Test set 4}  \\ 
\cline{2-9}
                        & DA             & w/o DA         & DA             & w/o DA         & DA             & w/o DA         & DA             & w/o DA          \\ 
\hline
\textbf{Ours}           & \textbf{0.827} & \textbf{0.810} & \textbf{0.783} & \textbf{0.743} & \textbf{0.731} & \textbf{0.617} & \textbf{0.802} & \textbf{0.654}  \\ 
\hline
U-Net~\cite{ronneberger2015u}                    & 0.823          & 0.817          & 0.769          & 0.737          & 0.712          & 0.639          & 0.781          & 0.670           \\ 
\hline
LinkNet~\cite{chaurasia2017linknet}                 & 0.724          & 0.720          & 0.665          & 0.632          & 0.608          & 0.451          & 0.663          & 0.491           \\ 
\hline
MNet~\cite{7950555}                    & 0.818          & 0.797          & 0.764          & 0.712          & 0.706          & 0.593          & 0.782          & 0.619           \\ 
\hline
Deeplabv3+~\cite{chen2018encoder}              & 0.814          & 0.778          & 0.737          & 0.656          & 0.681          & 0.448          & 0.731          & 0.431           \\ 
\hline
HRNet~\cite{wang2020deep}                   & 0.819          & 0.802          & 0.765          & 0.703          & 0.702          & 0.589          & 0.767          & 0.622           \\
\hline
\end{tabular}
\label{table1}
\end{table}

\subsection{Structure-wise Quantitative Comparison}

\begin{figure}[h]
\begin{center}
\begin{tabular}{c} 
\includegraphics[width=1\textwidth]{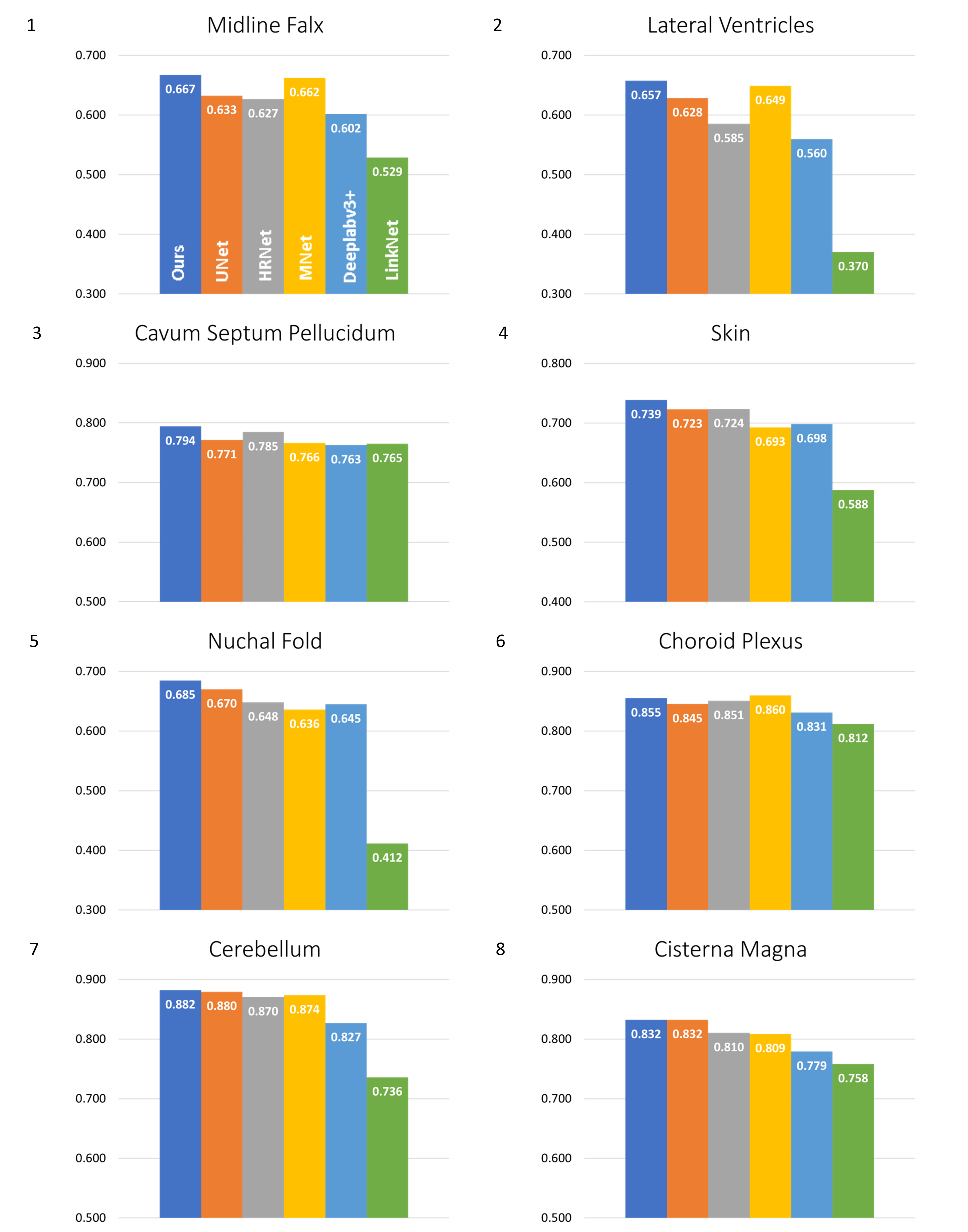}
\end{tabular}
\end{center}
\caption{\label{fig:structurecollage} 
The comparison of Dice-coefficients of individual structures (1: Midline Falx, 2: LV, 3: CSP, 4: Skin, 5: Nuchal Fold, 6: Choroid Plexus, 7: Cerebellum, 8: Cisterna Magna) averaged over the 4 test sets is presented.}
\end{figure} 

Further structure-wise analysis of model performance is provided in \textbf{Figure \ref{fig:structurecollage}}. The results of each structure are averaged across all the 4 test sets and presented in the figure. As it is evident from \textbf{Table \ref{table1}} and \textbf{Figure \ref{fig:structurecollage}}, our proposed approach generally performs better than benchmarking models. While the overall performance of U-Net model is similar (1.4 \% lower than our approach when averaged across all test sets), our approach outperforms U-Net specifically for finer structures like CSP, LV and nuchal fold which are crucial for differential diagnosis (DD) of CNS anomalies.

In fetal brain examination certain structures are more important. Changes in structures like deviation from normal CSP ratio and dilated lateral ventricles are small but crucial for the differential diagnosis of several CNS anomalies. Our proposed approach always outperforms the U-Net for these structures in all the test sets. Numerical values for CSP: Test set 1 (Proposed-0.863, U-Net-0.857), Test set 2 (Proposed-0.767, U-Net-0.739), Test set 3 (Proposed-0.836, U-Net-0.790), Test set 4 (Proposed-0.710, U-Net-0.698). Numerical values for LV: Test set 1 (Proposed-0.720, U-Net-0.706), Test set 2 (Proposed-0.612, U-Net-0.604), Test set 3 (Proposed-0.603, U-Net-0.573), Test set 4 (Proposed-0.693, U-Net-0.630).

\section{CONCLUSION}

In this study, we presented a custom DL framework for the automated 
segmentation of 10 axial view structures (TC and TV planes) of the fetal brain. On testing across 4 test sets, the proposed DL framework outperformed other popular segmentation networks by 0.5\% to 2\% (mean DC), localizing fine structures better. We have also presented the rationale behind selecting the feature extraction block based on empirical evidence of higher performance on important finer structures. Additionally, our proposed domain-specific DA improved model generalizability and showed significant performance improvements even across the benchmarking models. From training dataset ablation studies, we observed large segmentation performance gains and improved generalizability by simply using domain-specific DA, despite limited training data. 

We believe, our work opens doors for the development of device-independent and vastly generalizable DL models, a necessity for the seamless clinical translation and deployment. We hope the emergence of such assistive technology in clinics can truly help democratize quality and standardized prenatal care for every expecting mother. Our study faced limitations in terms of size of dataset, variation in demographics and number of brain examination planes and only considered normal fetal brain scans. Future works can focus on analyzing the specific performance differences across USG devices and wider demographic, use of domain adaptation, and performance evaluation on non-standard planes and anomaly cases. The hyperparameters for data augmentations have been set manually and future studies can also focus on automating the process and the dataset scales up.

\bibliography{report} 
\bibliographystyle{spiebib} 

\end{document}